# On the magnetic field of bound charges current.

*Alexander I. Korolev, Saint-Petersburg, Russian Federation*

*e-mail: alex-korolev@ya.ru*

Measurements of magnetic induction near an ice rod in the strong electric field were carried out. Theoretical estimation of the magnetic induction was made. It was found that in the average the experimental values of magnetic induction were an order less than theoretical value. A conclusion about the non- equivalence of magnetic fields of bound charges current and conduction current was made. Therefore, the magnetic field near an energized conductor is caused not directly by moving charges but by their influence on the propagation medium. The similar effect could occur at diffusion of other particles through the medium.

*March, 2013*

## Introduction

The bound charges current is a component of total current in the system of Maxwell's equations. Its dimension is the same as for the free charges current, namely, Ampere. As it is stated in classical electrodynamics, magnetic action of this current is similar to the action of conduction current [1]. The pioneer investigators of magnetic action of bound charges current were Roentgen [2] and Eichenwald [3,4]. The conducted experiments consisted in study of the magnetic field near the rotating disks of polarized dielectric. The discs were placed in electric field of flat conductive plates that caused induction of the surface bound charges. Sensitive magnetic needles were used to measure the magnetic field. Density of bound charges current at disc rotation was calculated:

$$J_b = b \cdot u \cdot \sigma_b \quad (1)$$

Here $\sigma_b$ is the surface density of bound charges, $u$ is the angular velocity, $b$ is a numerical coefficient. Roentgen discovered and qualitatively investigated the magnetic field of bound charges. Eichenwald carried out quantitative research, improving the setup and measurement technique. According to the obtained results, it was concluded that magnetic field of the bound charges current was identical to this one of the conduction current. However, the procedure of the magnetic field measurement using magnetized needles is not correct. It was the degree of influence of the charges on the existing magnetic field of the needle that was measured but not the magnetic field of moving charges directly. Sensors, operating on the Hall effect, could be used to measure the magnetic field correctly. The magnetic field of electron current in them is negligible.

## Measurement of magnetic field of bound charges current in ice

The current of bound charges appears when they move linearly and simultaneously as well as when electrical dipoles rotate around the center of mass. Essential difference between these types of currents is that in the former case, the current can be maintained continuously, and in the second case - only in the form of short pulses during the polarization establishing. The advantage of the second method is the essential current density, due to contribution of rotation of every electric dipole into the total current. This allows the use of Hall's sensors with average sensitivity to measure the magnetic field. Reasoning from their response time, ice with temperature of 0 ° C is chosen as a dielectric to be investigated. The ice is made by freezing distilled water.

Scheme of the experimental setup is presented in Fig. 1. High-voltage pulses are formed in a pulse generator ("Laska- super" shocker) and are fed onto a thin copper plate 100 x 100 mm in size. An isolated ceramic plate 8 mm in thickness is fixed on the copper plate.  A plastic cylinder with walls 0.5 mm thick, 10 mm in diameter and 20 mm in length is placed in the centre of the isolated plate. The cylinder is filled up with thawing ice at 0 °C. The magnetic field of current pulses from molecules $H_2O$ is measured by two Hall's probes Honeywell SS495A1. The sensors are located closely with wide faces to each other, at that their orientation is opposite with respect to the field. Sensitivity of the sensors under normal conditions is 3.1 mV / G, and the response time is 3 μs. The sensors are fixed normally to the imaginary magnetic field lines around the cylinder with ice (a piece of the conductor). Distance from the center of the sensor to the center of the cylinder is about 10 mm. To shield the sensors against the strong electric field from the copper plate, a grounded shield of aluminum foil on a plastic frame is used. A signal from the sensor is fed to the digital oscilloscope Hantek DSO-2090 and analyzed by PC. Ground connections of the oscilloscope and the screen are electrically isolated.

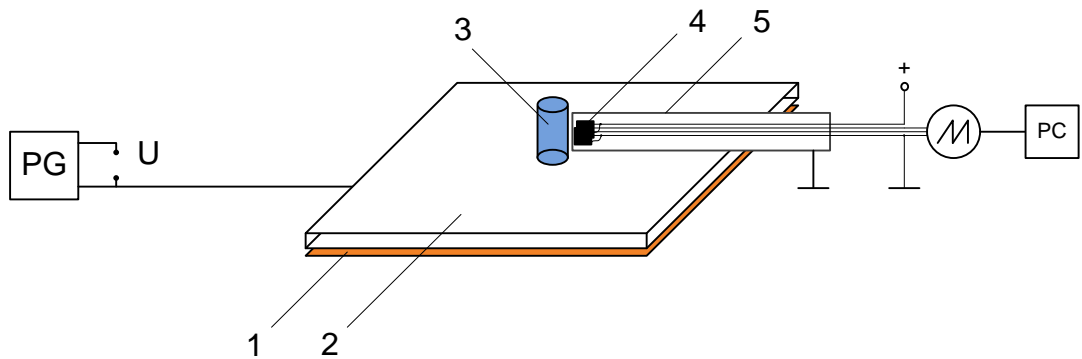

*Fig. 1. Scheme of experimental setup for measurement of ice rod magnetic field. Notes: 1 - copper plate, 2 - ceramic plate, 3 - plastic cylinder with ice, 4 - two Hall's sensors, 5 - electrostatic screen.*

Fig. 2 shows the voltage pulses U(t), incoming onto the copper plate from the pulse generator (PG). The pulse size is limited by a discharge gap at the output of the PG. The pulses are obtained by processing the electric signal from the piece of the conductor located near the plate. Normalization of the received signal is made to the nameplate output voltage of the PG. The electric field intensity above the center of the ceramic plate is:

$$E(t) \cong U(t)/2L \quad (2)$$

Here L is the distance from the center of the copper plate to the points of ice molecules location, $L \ll 100$ мм. The value of intensity in the area of the cylinder with ice during the voltage pulse is: $10^6 < E < 3{,}5 \cdot 10^6 \ V/m$. These values are significantly smaller than the value of intensity of water molecules decomposition. Thus, it is possible to carry out multiple repolarization with no destruction of the ice structure.

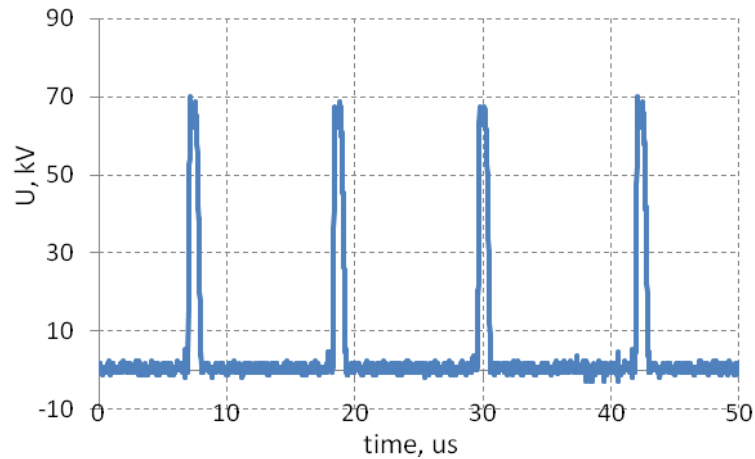

*Fig. 2. Shape of pulses incoming onto the copper plate from pulse generator.*

Two series of measurements are carried out, during which the signals from the magnetic sensors are recorded when applying voltage pulses. In the first series the cylinder with ice is present, in the second it is removed. Each series consists of two parts. One part includes 25 measurements. The series of measurements were divided into two parts in order to avoid significant melting of outer layers of the ice rod, placed under normal conditions (no more than 0.5 mm). In data processing the signals from magnetic sensors are subtracted to reduce noise and double the signal of interest. The differential signals received in this way are processed with a digital low-pass filter, the cutoff frequency is 1 MHz. Then we sum up the 50 signals in the series with the ice rod, as well as the 50 noise signals in the series without ice. After that, the total signals are subtracted. It enables to remove the systematic measurement error occurred due to the stray fields. The accuracy of the measurements of the magnetic induction is limited by ADC resolution of the oscilloscope, which makes up 0.2 mV (0.064 G for one magnetic sensor). The resulting voltage is converted to the value of magnetic induction corresponding to a measurement of one sensor by dividing by the number of measurements made by both sensors (100). Such method is applicable due to synchronization of the signals recording against the voltage pulses in course of the measurements. The result of signals processing from the Hall sensors is shown in Fig. 3.

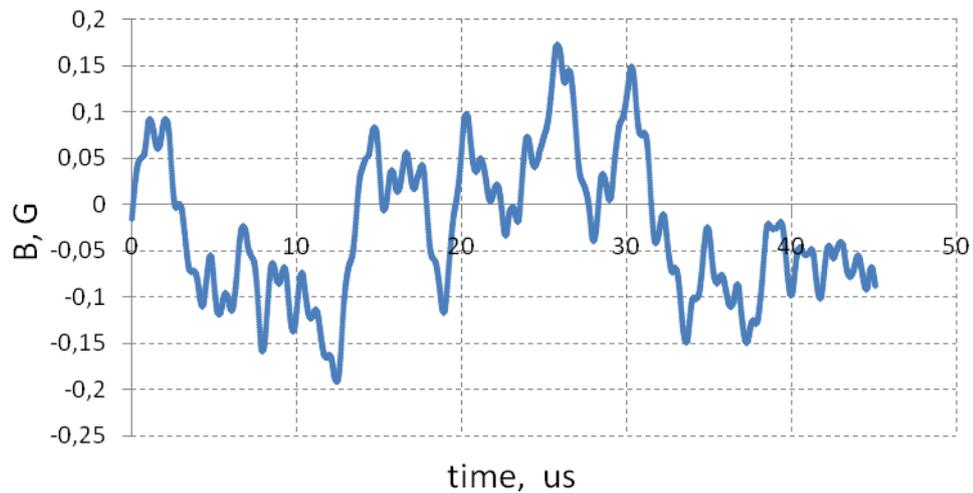

*Fig. 3. Dependence of magnetic induction on time near ice rod with the imposed electric field calculated from experimental data.*

In the obtained dependence there is no periodic signal with a frequency corresponding to the electric pulse-repetition interval. The dependence is like a noise, the mean deviation of the values of magnetic induction is 0,063 G.

## Theoretical estimation of magnetic field of bound charges current in ice

The bound charges current in process of ice polarization represents rotation of charges around the center of mass of their electric dipoles. At that the positive hydrogen ions shift in one direction and the negative oxygen ions - in the opposite. The shift occurs along the axis of the plastic cylinder. Let us investigate the shift process of the charges on micro-level to calculate the current flowing through the ice rod during the pulse propagation. The explanatory illustration to the calculation is given in Fig. 4.

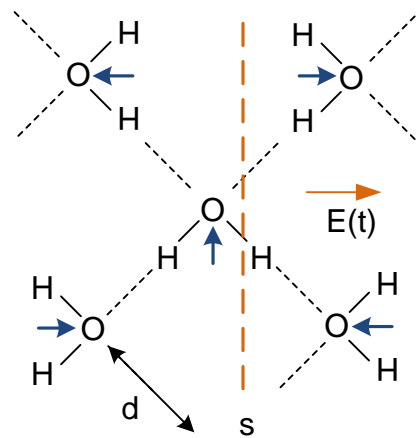

*Fig. 4. Illustration to estimation of the bound charges current in ice. $d \approx 2,76$ Å is the average interparticle distance , s is the cross- section of ice.*

In the absence of the external field the electric dipoles of water molecules are directed in different directions. When the electric field is imposed, the dipoles tend to turn along the direction of the field. Polarization relaxation time of ice at 0 ° C, when the dipoles rotate, is 20 μs [5-7]. Average distance between the water molecules in ice makes up 2,76 Å. Ice has a high viscosity that stabilizes the rotation process of the dipoles under the influence of the external electric force. Thus, we can assume that the dipoles rotation is uniform when an electric field is imposed. During the propagation of the electric field pulse ($t \approx 1,5$ μs) each dipole turns on average by *1.5 / 20 = 0.075* fraction of the full turn. At that, an average charge passing through a cross section of the rod, along the axis in both directions, makes up

$$Q = N \cdot (|q_+| + |q_-|) \quad (3)$$

Here N is the number of water molecules in the cross section, $q_+$ is the part of the charge of two hydrogen ions passing through the cross section at the point of time *t*, averaged over the cross section, $q_-$ is the part of the charge of one oxygen ion passing through the cross section at the point of time *t*, averaged over the cross section. Estimation of the number of molecules in the ice rod cross section gives: $N \sim \left(\frac{0.01}{2.76 \cdot 10^{-10}}\right)^2 \cong 1.3 \cdot 10^{15}$.

Let us simplify the task to estimate the charge that passes through the cross section of the ice rod in the polarization relaxation time. Let us consider the case of an "ideal" section passing through the centers of mass of *N* dipoles. In this case, when the dipoles are oriented towards the external field initially, the maximum transferred charge is the charge of all three ions and makes up *3e*. When the dipole is oriented initially along the field, the charge is equal to zero. The average transferred charge is equal to *3/2e*. Let us consider now the case of an "ideal" boundary section, separating the layers of N dipoles. Flow of the charge through it is absent. Ideally, when the molecules of ice are placed in the planes oriented perpendicular to the axis; those two types of sections are extreme. We can estimate the average charge that flows through the middle section between these extreme sections. It is approximately    . In a real crystal of ice the water molecules form a complex spatial ensemble, this gives an amendment to the estimation of the charge transferred across the section.

Now we can estimate $|q_+| + |q_-|$ over the time *t*. It makes up $0,075 \cdot \frac{3}{4} e = 0,0562 \, e$. Total current of bound charges through the ice rod over the pulse time and according to the definition is:

$$I = \frac{Q}{t} \sim 7.8 \, A \quad (4)$$

Let us calculate the hypothetical intensity of magnetic field created by this current in the area of the magnetic sensor location. The calculation scheme is presented in Fig. 5.

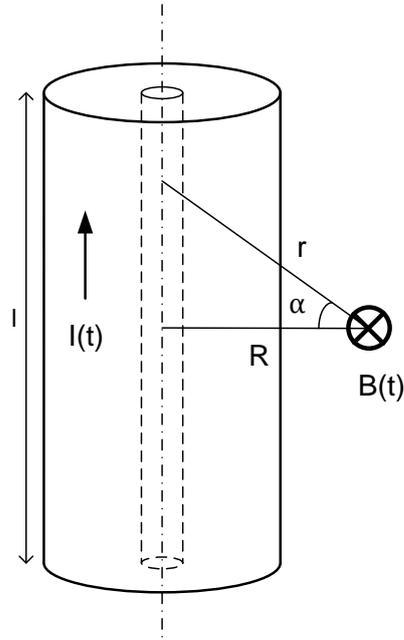

*Fig. 5. Calculation scheme for magnetic field generated by bound charges current in ice rod.*

To estimate the intensity we assume that all the current flows near the axis of the rod. Using the Biot-Savart law, we obtain an expression for the required magnetic induction of bound charges current in the segment of conductor (the ice rod):

$$B(t) \approx \frac{I(t)}{10^7} \cdot \int_0^{\text{arctg} \, l/2R} \frac{\cos^3 \alpha}{R} d\alpha \quad (5)$$

Substituting the known parameters into (5), we get an estimate of the magnetic induction magnitude at the point of location of the magnetic sensor over the pulse time: $B \sim 0{,}6$ G.

Thus, the magnitudes of the magnetic induction/noise (see Fig. 3) calculated from the experimental data in the average are an order less than the theoretical estimate.

## Results and discussion

The lack of magnetic field of the bound charges current in the ice is ascertained with accuracy of an order of magnitude of the magnetic induction. This leads to a conclusion about the nonequivalence of magnetic fields of bound charges current and conduction current. Thus, the magnetic field is not caused by moving electric charges only [1], but it is due to their influence on the propagation medium. Symmetry breakdown of electron shells that explains the magnetic properties of magnetoactive atoms may also cause magnetism in the magnetically inactive medium. Symmetry breakdown of the electron outer shells occurs at the directed flow of electrons through a substance.

Magnetic field of moving electrons was found in investigation of glow and arc discharges. The experiments were carried out under the supervision of A. F. Ioffe in the early 20th century [8]. However, magnetized needles were used to measure the magnetic field. Their deviations could be caused by the force that is opposite to the Lorentz force. The backward force should appear, according to Newton's third law, from the moving electrons and ions of the discharge. It also explained deviation of the needles in [2-4]. The further experience of scientists and engineers worked with electrovacuum equipment did not reveal the magnetic properties of charged beams in the absence of external magnetic field.

## Conclusions

From the results of investigation the hypothesis occurs that the magnetic properties of a substance can be caused by diffusion both of charged and neutral particles through it. The key condition is the directional impact on the electron shell structure in the atoms of the substance. Demonstration of the inverse Faraday effect [9] shows the possibility of magnetization under the influence of diffusion of polarized photons (laser beam). The next possible candidate for magnetization is the beam of polarized neutrons.

## Acknowledgements

The author thanks a radio engineer under the pseudonym of Varjag, whose keenness of observation gave impetus to the present work and Mr. E. Ryabchikov for helpful comments.